\documentclass[10pt,journal]{IEEEtran}

\usepackage{siunitx}
\usepackage{bm}
\usepackage{graphicx} 
\usepackage{float}
\usepackage{cite}
\usepackage{amsmath,amssymb}
\usepackage[outdir=./]{epstopdf}
\usepackage{afterpage}
\usepackage{optidef}
\usepackage{orcidlink}
\usepackage{soul}
\usepackage{tikz}
\usepackage{fontawesome5} 
\usetikzlibrary{calc}
\usepackage{pgfplots}
\usetikzlibrary{shapes.geometric, fit, positioning, backgrounds, arrows.meta}
\usepackage{multirow}
\usepackage{subcaption}
\usepackage{bbm}

\usepackage{amsmath,amsfonts}
\usepackage{amssymb}
\usepackage{amsthm}
\usepackage{array}
\usepackage{textcomp}
\usepackage{stfloats}
\usepackage{url}
\usepackage{verbatim}
\usepackage{graphicx}
\usepackage{siunitx}

\usepackage{mathtools}
\usepackage{dsfont}
\usepackage{bm}
\usepackage{mathrsfs}
\usepackage{enumitem}
\usepackage{xpatch}
\usepackage{caption}
\usepackage{subcaption}
\usepackage{adjustbox}
\usepackage[ruled,norelsize, noline]{algorithm2e}
\usepackage{soul}
\usepackage{tikz}
\usepackage{tikzscale}
\usetikzlibrary{calc}
\usetikzlibrary{arrows, positioning}
\usetikzlibrary{shapes}
\usetikzlibrary{pgfplots.groupplots}
\usetikzlibrary{spy}
\usepackage{pgfplots}
\usepackage{scalefnt}
\usetikzlibrary{arrows.meta}
\usepackage[font=small]{caption}
\usepackage{pgfplots}
\usepgfplotslibrary{groupplots}
\usepackage{xcolor}
\usepackage{booktabs}

\usepackage[nolist]{acronym}

\usepackage{cite}
\usepackage{color}

\newacro{ATC-FWI}[ATC-FWI]{Adapt-Then-Combine Full Waveform Inversion}
\newacro{FWI}[FWI]{Full Waveform Inversion}
\newacro{NN}[NN]{Neural Network}
\newacro{ML}[ML]{Machine Learning}
\newacro{JSCC}[JSCC]{Joint Source and Channel Coding}
\newacro{FL}[FL]{Federated Learning}
\newacro{OAC}[AirComp]{Over-the-Air Computation}
\newacro{ELBO}[ELBO]{Evidence Lower BOund}
\newacro{MILBO}[MILBO]{Mutual Information Lower BOund}
\newacro{dELBO}[dELBO]{Distributed Evidence Lower BOund}
\newacro{dMILBO}[dMILBO]{Distributed Mutual Information Lower BOund}
\newacro{AWGN}[AWGN]{Additive White Gaussian Noise}
\newacro{DNN}[DNN]{Deep Neural Network}
\newacro{SGD}[SGD]{Stochastic Gradient Descent}
\newacro{MSE}[MSE]{Mean Square Error}
\newacro{MAS}[MAS]{multi-agent system}  
\newacro{CNN}[CNN]{Convolutional Neural Network}
\newacro{NMSE}[NMSE]{Normalized Mean Square Error}
\newacro{SSIM}[SSIM]{Structural Similarity Index Measure}
\newacro{ViT}[ViT]{Vision Transformer}
\newacro{ResNet}[ResNet]{Residual Network}
\newacro{PReLU}[PReLU]{Parametric Rectified Linear Unit}
\newacro{MSE}[MSE]{Mean Square Error}
\newacro{NN}[NN]{Neural Network}
\newacro{PSNR}[PSNR]{Peak Signal to Noise Ratio}
\newacro{SNR}[SNR]{Signal to Noise Ratio}
\newcommand{\s}{s}
\newacro{pdf}[PDF]{Probability Density Function}

\newcommand{\weightSSIM}{\ensuremath{w_{i_n}}}  
\newcommand{\expct}{\ensuremath{\mathrm{E}}}
\newcommand{\entropy}{\ensuremath{\mathrm{H}}}

\DeclareMathOperator*{\argmin}{argmin}
\DeclareMathOperator*{\argmax}{argmax}

\DeclareMathAlphabet\mathboldsf{OT1}{cmss}{bx}{n}

\newlength\figH
\newlength\figW
\newlength\figWsmall
\newcommand*{\nth}[2]{
    \foreach \xindx [count=\kcount from 0] in #1 {
        \ifnum\kcount=#2
            \xindx
        \fi
    }
}

\newboolean{xtickAutomatic}
\setboolean{xtickAutomatic}{true}

\newboolean{ytickAutomatic}
\setboolean{ytickAutomatic}{true}

\newboolean{xMinAutomatic}
\setboolean{xMinAutomatic}{true}
\newboolean{xMaxAutomatic}
\setboolean{xMaxAutomatic}{true}

\newboolean{yMinAutomatic}
\setboolean{yMinAutomatic}{true}
\newboolean{yMaxAutomatic}
\setboolean{yMaxAutomatic}{true}

\newboolean{LegendAutomatic}
\setboolean{LegendAutomatic}{true}

\usepackage[absolute,overlay]{textpos}

\IEEEoverridecommandlockouts 

\usepackage{hyperref}  
\hypersetup{
	colorlinks=false,
	linkbordercolor=white,
	urlbordercolor=white,
	pdfborder={0 0 0}
}  
\usepackage{adjustbox}

\begin{document}

\title{Semantic Communication for Task Execution and Data Reconstruction in Multi-View Scenarios} %


\author{Maximilian H. V. Tillmann\,\orcidlink{0009-0000-6548-4278},~\IEEEmembership{Graduate Student Member,~IEEE,}
        Avinash Kankari\,\orcidlink{0009-0002-2947-6945},   \\     
         Carsten Bockelmann\,\orcidlink{0000-0002-8501-7324},~\IEEEmembership{Member,~IEEE,}
        Armin Dekorsy\,\orcidlink{0000-0002-5790-1470},~\IEEEmembership{Senior Member,~IEEE}
\thanks{This work was partially supported by the Deutsche Forschungsgemeinschaft (DFG, German Research Foundation) under Germany's Excellence \mbox{Strategy – EXC-3036} The Martian Mindset, project number: 533607631,
by the DFG under project number 500260669 (SCIL), and by
the German Ministry of Research, Technology and Space (BMFTR) under project number 16KISK016 (Open6GGub).
The authors are with the Department of Communications Engineering,
University of Bremen, 28359 Bremen, Germany. 
E-mails: \{tillmann, bockelmann, dekorsy\}@ant.uni-bremen.de
}%
}

\maketitle


\begin{textblock*}{2\columnwidth}(1.9cm,26.8cm)
\centering
\footnotesize
DOI: 10.1109/LWC.2026.3712239 ~\copyright~2026 IEEE \url{https://ieeexplore.ieee.org/abstract/document/11603385}
\end{textblock*}

\setlength{\footskip}{18pt}

\begin{abstract}

Semantic communication has gained significant attention with the advances in machine learning. 
Most semantic communication works focus on either task execution or data reconstruction, with some recent works combining the two.
In this work, we propose a semantic communication system for concurrent task execution and data reconstruction for a multi-view scenario, which we formulate as the maximization of mutual information. To investigate the trade-off between the two objectives, we formulate a joint objective as a convex combination of task execution and data reconstruction.
We show that under specific assumptions, the \ac{SSIM} loss can be obtained from the mutual information maximization objective for data reconstruction, which takes human visual perception into account.
Furthermore, for constant resource use, we show that by increasing the weight of the reconstruction objective up to a certain point, the task execution performance can be kept nearly constant, while the data reconstruction can be significantly improved.


\end{abstract}

\setlength{\jot}{2pt} 
\setlength{\belowdisplayskip}{4pt}
\setlength{\belowdisplayshortskip}{4pt}
\setlength{\abovedisplayskip}{4pt}
\setlength{\abovedisplayshortskip}{4pt}


\section{Introduction} \label{section.Intro}
\IEEEPARstart{W}{ith} the recent advances of deep learning methods, semantic communication is expected to play a key role in the development of future communication systems of 6G and \mbox{beyond}  \cite{Gunduz2022, getu2025semantic,WenTong}.
Unlike Shannon's goal of accurately transmitting every bit of message, semantic communication has the goal of accurately conveying the intended meaning behind a message \cite{Sana2022}.

Most works on semantic communication can be grouped into one of the two objectives:
(i) Task execution, where a single or multiple agents observe some data, with the goal to execute a task at a remote location, meaning that all observation data is not required to be transmitted \cite{Gunduz2022,Shao2021,Beck2023}. 
(ii) Data reconstruction, where the goal is to fully reconstruct all data, e.g., text, speech, or image data, according to a semantic accuracy measure \cite{Xie2021,Gunduz2022}.

\begin{figure}[t!]
\centering
    \resizebox{0.49\textwidth}{!}{ 
    \includegraphics{tikz/block_diagram/distributed_task_execution.tikz}
    }
    \vspace{-0.6cm}
    \caption{
    Multi-view semantic communication with $N$ sensing nodes transmitting observations to a receiver for joint task execution (e.g., fault detection) and data reconstruction.
    \vspace{-1.5em}  
    }
    \label{fig:scenario}
\end{figure}

However, in many scenarios, it might be advantageous if both task execution and data reconstruction are possible. One such example is shown in Fig.~\ref{fig:scenario} with error detection in a production line. If an error is missed and only identified at the end of the production line, data reconstruction is required so that a human expert can review the sensor data to localize the fault.
Furthermore, if the task changes or new data processing methods are to be applied to the original data in the future, data reconstruction alongside task execution is required if retransmission is not feasible or not desired.

Such scenarios of joint task execution and data reconstruction were only recently investigated in \cite{lyu2024semantic} and \cite{yuan2024generative}, where it was shown that if the communication system is only optimized for task execution, reconstruction is only possible in a limited way, and vice versa. 

In \cite{lyu2024semantic} the authors focused on the scenario of image classification and image reconstruction under training data label corruption. They formulated a coding rate reduction maximization problem to achieve robustness to training data label corruption, and used the \ac{MSE} loss for the image reconstruction. Other loss functions for image reconstruction, like \acf{SSIM}, which is a measure of image similarity aligned for human visual perception \cite{hitchhikersguideSSIM}, were not considered in their proposed approach. A system with \ac{SSIM} as the loss function served only as a baseline, and the proposed methods were not evaluated for \ac{SSIM}.
In \cite{yuan2024generative}, the authors investigated joint image reconstruction and image segmentation. They used the cross entropy loss for image segmentation and the \ac{MSE} loss for image reconstruction, again without considering other loss functions for image reconstruction like \ac{SSIM}.

    
In contrast to these existing works, we focus on extending the image reconstruction problem beyond the \ac{MSE} loss.
To this end, we propose a unified framework for semantic communication system design, in which both image reconstruction and image classification are formulated as a mutual information maximization problem.
We show how different image reconstruction losses (\ac{MSE}, \ac{SSIM}, or a combination of both) can be derived from the mutual information maximization problem under different assumptions on the decoder \ac{pdf}.
Furthermore, we evaluate the trade-off between image reconstruction quality, and task execution error rate for our proposed method.
    
Finally, we mention that in contrast to the existing methods of joint task execution and data reconstruction, we consider the distributed multi-view scenario for classification. 
Each sensing node observes only part of the data, requiring extraction of task-relevant semantics for classification at the decoder. Unlike the centralized case, optimal classification cannot be achieved by any single sensing node alone as information from other views may be required.

Based on the above discussions, the main contributions are summarized as follows:
\begin{itemize}[left=0pt]
    \item We propose a semantic communication system for joint task execution and data reconstruction in multi-view scenarios, formulated as a mutual information maximization problem.
    
    \item We show the derivation of \ac{SSIM} loss from the mutual information maximization problem.


    \item We analyze the trade-off between task execution and data reconstruction. 
    Increasing the reconstruction weight improves reconstruction with little impact on task performance, but overly weighting on reconstruction degrades task execution.
\end{itemize}

\section{System Model and Problem Formulation} \label{section.SystemModel}

\subsection{Semantic Communication System Model}
Our distributed semantic communication system model is shown in Fig.~\ref{fig:system_model}. 
We assume a semantic source as the joint probability distribution $p(\bm{\s}_1,\dots,\bm{\s}_{N},\bm{z})$ of the observations $\bm{\s}_1,\dots,\bm{\s}_{N}$ and the semantic variable $\bm{z}$. 
The goal at the receiver is to reconstruct both the semantic variable, as well as all observations, which are denoted by $\hat{\bm{z}}$ and $\hat{\bm{s}} = \begin{bmatrix}
    \hat{\bm{s}}_1^\top ,\dots, \hat{\bm{s}}_N^\top
\end{bmatrix}^{\!\top}$, respectively.
We use one encoder per sensing node to optimally use the channel resources for the joint objective of task execution and data reconstruction. On the receiver side, we use separate decoders for the two objectives, as a design where the decoder first reconstructs the data and then uses this data for task execution cannot improve performance due to the data processing inequality \cite{cover1999elements}.

%


\subsection{Problem Formulation for Joint Task Execution and Data Reconstruction}
We formulate our optimization problem as the convex combination of two mutual information terms that should be maximized to balance the trade-off between the two objectives. 
We design the encoders $p_{\bm{\theta}_i}(\bm{c}_i|\bm{\s}_i)$ with \ac{NN} parameters $\bm{\theta}_i$, such that the received symbols $\bm{y}$ are  most informative about the semantic variable $\bm{z}$ for task execution with weight $1-\alpha$, and about the observations $\bm{s}$ for data reconstruction with weight $\alpha\in[0,1]$. 
As shown in Fig.~\ref{fig:system_model}, $\bm{y}$ depends on $\bm{\theta}_1,\ldots,\bm{\theta}_N$ via $\bm{y}\sim p(\bm{y}|\bm{c}_1,\ldots,\bm{c}_N)$ and $\bm{c}_i\sim p_{\bm{\theta}_i}(\bm{c}_i|\bm{s}_i)$. For simplicity, we do not make this dependency explicit in our notation.
We optimize the \ac{NN} parameters $\bm{\theta}_i$ of the 
encoders distributions $p_{\bm{\theta}_i}(\bm{c}_i|\bm{\s}_i)$ as
\vadjust{\allowbreak}
\begin{align} \label{InfoMax}
    &\argmax_{\bm{\theta}_1,\dots,\bm{\theta}_N } \ \  \alpha I(\bm{s};\bm{y}) + (1-\alpha)I(\bm{z};\bm{y}) \\
    & \quad \text{s.t.} \ \ \bm{c}_i  \in \mathbb{R}^{N_{\text{Tx}}/N} , \, P_i \leq 1 \nonumber, \ \text{for} \ i=1,\dots,N ,
\end{align}
where ${\bm{s}} = \begin{bmatrix}
    {\bm{s}}_1^\top,\dots,{\bm{s}}_N^\top
\end{bmatrix}^\top$, $N_{\text{Tx}}$ is the number of total channel 
uses with $\bm{y}\in\mathbb{R}^{N_{\text{Tx}}}$, and $P_i= 1/(N_{\text{Tx}}/N) \sum_{n=1}^{N_{\text{Tx}}/N}{c}_{i_n}^2$
is the power constraint for node $i$ per channel use,
where ${c}_{i_n}$ is the $n$-th element of $\bm{c}_i$. The mutual information between $\bm{z}$ and $\bm{y}$ \cite{cover1999elements} is given as $I(\bm{z};\bm{y})= \expct_{p(\bm{z},\bm{y})}\left[\log \left( \frac{p(\bm{z} | \bm{y})}{p(\bm{z})}\right)\right]$,
where $\expct_{p(x)}[\cdot]$ is the expectation with respect to $x\sim p(x)$, and $\log(\cdot)$ is the natural logarithm.
However, the posterior $p(\bm{z} | \bm{y})$ is typically not available, making it infeasible to optimize the encoder for a $\bm{y}$ that maximizes the mutual information terms \cite{Beck2023}. A common approach is to use variational approximations of the posterior, $q_{\bm{\psi}}(\bm{z} | \bm{y})$ with parameters $\bm{\psi}$ and $q_{\bm{\phi}}(\bm{s} | \bm{y})$ with parameters $\bm{\phi}$ for the semantic decoder and the reconstruction decoder, respectively. 
This way, a lower bound of the mutual information can be derived that can be optimized instead.
The following derivations are only shown for $I(\bm{z};\bm{y})$, but can be analogously obtained for $I(\bm{s};\bm{y})$ by replacing $\bm{z}$ by $\bm{s}$.
\begin{align}
 \! \! I(\bm{z};\bm{y}) 
                            & = \expct_{p(\bm{z},\bm{y} )}\left[ \log\left( \frac{p(\bm{z}|\bm{y} )}{p(\bm{z})} \frac{q_{\bm{\psi}}(\bm{z}|\bm{y} )}{q_{\bm{\psi}}(\bm{z}|\bm{y} )}\right) \right]\\
                            & = \expct_{p(\bm{y} )}\left[ \expct_{p(\bm{z}|\bm{y} )}\left[ \log\left( q_{\bm{\psi}}(\bm{z}|\bm{y} )\right) \right] \right] \nonumber \\
                            & \ \ \  - \expct_{p(\bm{z})} \!\left[  \log\left( p(\bm{z})\right) \right] + \expct_{p(\bm{z},\bm{y} )} \! \!\left[ \log \!\left( \!\frac{p(\bm{z}|\bm{y} ) }{q_{\bm{\psi}}(\bm{z}|\bm{y} )} \!\right) \! \right]  \\
                            & = - \expct_{p(\bm{y} )}\left[\entropy\left( p(\bm{z}|\bm{y} ), q_{\bm{\psi}}(\bm{z}|\bm{y} )\right) \right] \label{condMI_identity} \nonumber \\
                            & \ \ \ + \entropy(\bm{z})   + \expct_{p(\bm{y})}\left[D_{\text{KL}}\left( p(\bm{z}|\bm{y}) || q_{\bm{\psi}}(\bm{z}|\bm{y} ) \right) \right] \\
                            &\geq - \expct_{p(\bm{y} )}\left[\entropy\left( p(\bm{z}|\bm{y} ), q_{\bm{\psi}}(\bm{z}|\bm{y} )\right)\right]  + \entropy(\bm{z})  \label{MILBo_short},
\end{align}
where $\entropy(p,q)$ is the cross-entropy between $p$ and $q$, $\entropy(\cdot)$ is the entropy, and $D_\text{KL}(p,q)$ is the Kullback-Leibler (KL) divergence from $q$ to $p$. The last step uses that the KL divergence is non-negative \cite{cover1999elements}. From \eqref{condMI_identity} and \eqref{MILBo_short}, it can be seen that the lower bound on the mutual information is tight if the KL divergence is zero, meaning that the true posterior 
$p(\bm{z}|\bm{y} )$ and its variational approximation $ q_{\bm{\psi}}(\bm{z}|\bm{y} )$ are identical almost everywhere \cite{cover1999elements}.

\begin{figure}[t!]
\centering
    \resizebox{0.50\textwidth}{!}{
    \includegraphics{tikz/block_diagram/system_model.tikz}
    }
    \vspace{-1.5em}  
    \caption{Multi-view semantic communication system model for joint task execution and data reconstruction. 
    }
    \vspace{-1.8em}  
    \label{fig:system_model}
\end{figure}

The objective in \eqref{InfoMax} now becomes the minimization of the mutual information lower bound \eqref{MILBo_short}. With the respective weights of $\alpha$ and $1-\alpha$ we have
\begin{align}
    & -\alpha \expct_{p(\bm{y})}\left[\entropy\left( p(\bm{s}|\bm{y} ), q_{\bm{\phi}}(\bm{s}|\bm{y} )\right)\right] +\alpha \entropy(\bm{s}) \\
    & - (1-\alpha) \expct_{p(\bm{y})}\left[\entropy\left( p(\bm{z}|\bm{y} ), q_{\bm{\psi}}(\bm{z}|\bm{y} )\right)\right] +(1-\alpha) \entropy(\bm{z})\nonumber,
\end{align}
which, by applying \eqref{condMI_identity}, can be rewritten as
\vadjust{\allowbreak}
\begin{align}  
    &\underbrace{\alpha I(\bm{s};\bm{y} ) + (1-\alpha)I(\bm{z};\bm{y})}_\text{weighted encoder objective}  - \alpha \underbrace{ \expct_{p(\bm{y})}\left[D_{\text{KL}}\left( p(\bm{s}|\bm{y} ) || q_{\bm{\phi}}(\bm{s}|\bm{y} ) \right)\right]}_\text{Reconstruction decoder objective} \nonumber \\
    &- (1-\alpha) \underbrace{ \expct_{p(\bm{y})}\left[D_{\text{KL}}\left( p(\bm{z}|\bm{y} ) || q_{\bm{\psi}}(\bm{z}|\bm{y} ) \right)\right]}_\text{Semantic decoder objective}, \label{enc_dec_objective}
\end{align}
which shows that minimizing the weighted mutual information lower bound implicitly optimizes encoder and decoder objectives. 
The objective for the encoders is to maximize the weighted mutual information for the reconstruction and the task execution. 
The objectives for the two decoders are minimizing the KL divergence between the true decoder and the variational approximation, which are independent of $\alpha$, as $q_{\bm{\phi}}(\bm{s}|\bm{y} )$ and $q_{\bm{\psi}}(\bm{z}|\bm{y} )$ are independent.
Finally, as the entropies $\entropy(\bm{z})$ and $\entropy(\bm{s})$ do not depend on any encoder or decoder parameters, our optimization problem is the minimization of the weighted cross-entropy terms with respect to the \ac{NN} parameters $\{\bm{\theta}_1,\dots,\bm{\theta}_N, \bm{\phi}, \bm{\psi}\}$ given as
\vspace{-0.4cm}
\begin{align} \label{MinCrossEntropy}
    & \underset{\substack{\bm{\theta}_1,\dots,\bm{\theta}_N, \bm{\phi} , \bm{\psi} }}{\argmin} 
    \begin{array}{l}
    \\ \ \
    \alpha \expct_{p(\bm{y})}\left[\entropy\left( p(\bm{s}|\bm{y} ), q_{\bm{\phi}}(\bm{s}|\bm{y} )\right)\right]\\+ (1-\alpha)\expct_{p(\bm{y})}\left[\entropy\left( p(\bm{z}|\bm{y} ), q_{\bm{\psi}}(\bm{z}|\bm{y} )\right)\right] \end{array}\\
    & \quad \text{s.t.} \ \ \bm{c}_i  \in \mathbb{R}^{N_{\text{Tx}}/N} , \, P_i \leq 1, \ \text{for} \ i=1,\dots,N . \nonumber 
\end{align}

\vspace{-0.2cm}
\section{Solving the Optimization Problem} \label{sec:implementationConsiderations}
For our goal of solving \eqref{MinCrossEntropy}, we iteratively update the \ac{NN} parameters $\{\bm{\theta}_1,\dots,\bm{\theta}_N, \bm{\phi}, \bm{\psi}\}$ using stochastic gradient descent based optimization with the reparametrization trick \cite{Beck2023,GoodfellowDeepLearning}. 
The gradients of the cross-entropy terms of the optimization objective are obtained from the training data samples using Monte-Carlo approximation.

To optimize the cross-entropy of the reconstruction decoder $q_{\bm{\phi}}(\bm{s}|\bm{y})$ for image data, we have the problem of high dimensionality. Even for small color images of size $32\times32\times3$ with $8$ bit resolution, e.g., as in the CIFAR-10 dataset \cite{Krizhevsky09learningmultiple}, this means that there are $2^{8 \cdot 32\cdot32\cdot3}$ discrete probability events of $q_{\bm{\phi}}(\bm{s}|\bm{y})$, which are infeasible to estimate.
Therefore, $q_{\bm{\phi}}(\bm{s}|\bm{y} ) $ needs to be simplified. 
A simple way to reduce the complexity is to approximate the discrete $q_{\bm{\phi}}(\bm{s}|\bm{y} )$ by a parameterized continuous distribution with a tractable number of parameters.

In the following, we assume two different parameterized continuous distributions. We show how the \ac{MSE} and \ac{SSIM} losses are derived from the cross-entropy under the assumed distributions $q_{\bm{\phi}}^\prime(\bm{s}|\bm{y})$ and $q_{\bm{\phi}}^{\prime\prime}(\bm{s}|\bm{y})$, respectively.

\subsection{Deriving the MSE Loss} \label{sec:MSE}
We show that the \ac{MSE} loss is derived from the cross-entropy 
under the assumption that all $s_1,\dots,s_L$ are independently
Gaussian distributed with fixed and equal variances, where $s_n$ is the $n$-th element of $\bm{s}\in \mathbb{R}^L$\cite{GoodfellowDeepLearning}.



We call this distribution $q_{\bm{\phi}}^\prime(\bm{s}|\bm{y})$, which we model with a \ac{NN} with \ac{NN} parameters $\bm{\phi}$. The input to the \ac{NN} is $\bm{y}$ and the output is $\bm{\mu}(\bm{y})$, which depends on $\bm{y}$, and where $\mu_{n}(\bm{y})$ is the $n$-th element of $\bm{\mu}(\bm{y})\in \mathbb{R}^L$. In our notation, we omit the dependence of $\bm{\mu}(\bm{y})$ from the \ac{NN} parameters $\bm{\phi}$ for simplicity. 

With the above assumptions and a given variance $\sigma^2$, $q_{\bm{\phi}}^\prime(\bm{s}|\bm{y})$ is fully described by the mean $\bm{\mu}(\bm{y})$ of the Gaussian distribution. 
The cross-entropy then simplifies to
\vadjust{\allowbreak}
\begin{align}    
      &-\expct_{p(\bm{y})} \!\left[ \expct_{p(\bm{s}|\bm{y})}\left[\log\left( q_{\bm{\phi}}^\prime(\bm{s}|\bm{y} )\right) \right] \right] \nonumber \\ 
      & \!= \!- \! \expct_{p(\bm{y})} \! \!\left[ \!\expct_{p(\bm{s}|\bm{y})} \! \!\left[ \! \log \!\left( \prod_{n=1}^L \!  \frac{1}{\sqrt{2\pi\sigma^2}} \exp \! \!\left(\!\! -\frac{ \! \!\left( s_n - \mu_{n}(\bm{y})\right)^2 \!}{2\sigma^2 } \right) \! \! \right) \! \right] \right]\\[-0.5em] 
    &\! = \! \expct_{p(\bm{y})} \! \!\left[ \! \expct_{p(\bm{s}|\bm{y})} \! \!\left[ \! \sum_{n=1}^L \! \left(  \!\frac{\left(s_n - \mu_{n}(\bm{y})\right)^2}{2\sigma^2} \! \right) \! \right] \right] \! \! + \!  \frac{L}{2} \log \!\left( 2\pi\sigma^2\right) \!. \raisetag{10pt} \label{MSE_derived}
\end{align}

To minimize \eqref{MSE_derived} with fixed variance $\sigma^2$, we have to minimize $\expct_{p(\bm{y})}  \left[  \expct_{p(\bm{s}|\bm{y})} \left[ \text{MSE}(\bm{s},\bm{\mu}(\bm{y}))\right] \right]$, with \mbox{$\text{MSE}(\bm{s},\bm{\mu}(\bm{y}))=\frac{1}{L}\sum_{n=1}^L \! \left(s_n - \mu_{n}(\bm{y})\right)^2$}, which we approximate using the training data samples \cite{GoodfellowDeepLearning}.
Finally, using the maximum likelihood criterion to get the estimate $\hat{\bm{s}}$, we have $\hat{\bm{s}}=\bm{\mu}(\bm{y})$, as we assumed $q^\prime_{\bm{\phi}}(\bm{s}|\bm{y} )$ to be Gaussian, whose maximum occurs at $\bm{\mu}(\bm{y})$.


\subsection{Deriving the SSIM Loss} \label{sec:SSIM}
We show that the \ac{SSIM} loss is derived from the cross-entropy by assuming a certain parameterized distribution for the decoder.
The \ac{SSIM} between two images is defined as a weighted average over $M$ windows, where the weight associated with the $n$-th pixel of window $i$ is denoted by $\weightSSIM \!$ with $\sum_{n=1}^L \!\weightSSIM \! \!=  \!1$ for $i  \! =\! 1, \!{\tiny \dots},\!M$. Typically, these weights are specified by two dimensional finite-extent Gaussian or rectangular shaped windows \cite{hitchhikersguideSSIM}.
The \ac{SSIM} between images $\bm{s}$ and $\bm{\gamma}$ is then given as
\vspace{-0.1cm}
\begin{equation}
    \text{SSIM}(\bm{s},\bm{\gamma}) = \frac{1}{M} \scalebox{0.8}{$\displaystyle\sum_{i=1}^M$}    l_{i}(\bm{s},\bm{\gamma}) \, g_{i}(\bm{s},\bm{\gamma}), \label{ssim}
\end{equation}
\begin{equation}
    l_{i}(\bm{s},\bm{\gamma}) = \frac{2  \left(\sum_{n=1}^L w_{n_{i}}\gamma_{n}\right) \left(\sum_{n=1}^L w_{n_{i}}s_{n}\right)+c_1}{\left(\sum_{n=1}^L w_{n_{i}}\gamma_{n} \right)^2+\left(\sum_{n=1}^L w_{n_{i}}s_{n} \right)^2+c_1},
\end{equation}
\begin{align}
    g_{i}(\bm{s},\bm{\gamma}) \! = \! \frac{ \! \! \! \! \! \!  \! \!  \! \! 2 \displaystyle \scalebox{0.8}{$\displaystyle\sum_{n=1}^L$} w_{n_{i}} \! \!\gamma_{n}s_{n} \! - \! \left( \! { \scalebox{0.8}{$\displaystyle\sum_{n=1}^L$} } \weightSSIM \gamma_{n} \!\right) \! \! \! \left( \! \scalebox{0.8}{$\displaystyle\sum_{n=1}^L$} \weightSSIM s_{n} \!\right) \! \!  + \! c_2}{\displaystyle  \scalebox{0.8}{$\displaystyle\sum_{n=1}^L$} \!\weightSSIM \! \left(\gamma_{n}^2 \!+ \!s_{n}^{2} \right) \!- \!\! \left( \!\scalebox{0.8}{$\displaystyle\sum_{n=1}^L$} \weightSSIM \gamma_{n} \! \right)^{\! \! 2} \! \!  - \! \!\left( \! \scalebox{0.8}{$\displaystyle\sum_{n=1}^L$} \weightSSIM s_{n} \!\right)^{\! \!2} \!\!+ \!c_2 \!}, 
    \raisetag{11.5ex} 
\end{align} 
with constants $c_1, c_2 >0$. We have that $\text{SSIM}(\bm{s},\bm{\gamma})=1$ if and only if the images are identical 
\cite{hitchhikersguideSSIM}. Furthermore, $\text{SSIM}(\bm{s},\bm{\gamma})$ of $0$ indicates no structural similarity, and of $-1$ indicates negative correlation between the images \cite{hitchhikersguideSSIM}.

We define $q_{\bm{\phi}}^{\prime \prime}(\bm{s}|\bm{y})$, parameterized by $\bm{\gamma}(\bm{y})$, such that we get the \ac{SSIM} loss to minimize the cross-entropy. 
We model $q_{\bm{\phi}}^{\prime \prime}(\bm{s}|\bm{y})$ with a \ac{NN} with \ac{NN} parameters $\bm{\phi}$. The input to the \ac{NN} is $\bm{y}$ and the output is $\bm{\gamma}(\bm{y})$, which are the parameters fully describing $q_{\bm{\phi}}^{\prime \prime}(\bm{s}|\bm{y})$.
Again, in our notation, we omit the dependence of $\bm{\gamma}(\bm{y})$ from the \ac{NN} parameters $\bm{\phi}$ for simplicity. 
We normalized all pixel values to be in the range of $[0,1]$, 
and we define 
\vspace{-0.1cm}
\begin{equation} \label{decoder_SSIM}
\begin{aligned}
q_{\bm{\phi}}^{\prime \prime}(\bm{s}|\bm{y}) \!=\!
\begin{cases}
    \exp \! \left( \text{SSIM}\left(\bm{s}, \!\bm{\gamma}(\bm{y}) \right) - 1\right), & \! \!  \text{if} \ \bm{s}\in [0,d]^L,\\
    0, & \! \! \text{otherwise},
\end{cases} \! \! \!
\end{aligned}
\raisetag{2.5ex} 
\end{equation}
and $d$ such that $ \! \int_{0}^d \!... \!\int_{0}^d \! \exp   \!\left(\text{SSIM}\left(\bm{s}, \!\bm{\gamma}(\bm{y}) \right)  - 1\right)  ds_1  ...  ds_L \! \! = \! 1$.

As we have $\left| \text{SSIM}\left(\bm{s}, \!\bm{\gamma}(\bm{y}) \right) \right| \leq 1$ \cite{hitchhikersguideSSIM},  we have $e^{-2}\leq \exp \! \left( \text{SSIM}\left(\bm{s}, \!\bm{\gamma}(\bm{y}) \right) - 1\right) \leq 1$. Therefore, we have $d\in [1,e^{\frac{2}{L}}]$, such that $q_{\bm{\phi}}^{\prime \prime}(\bm{s}|\bm{y})$ is not zero for $\bm{s}\in [0,1]$, which is the rage the image data is normalized to.
We have shown now that $q_{\bm{\phi}}^{\prime \prime}(\bm{s}|\bm{y})$ is a \ac{pdf}, since is non-negative and integrates to one over its domain.

Now, for $\bm{s} \in [0,d]^L$, the cross-entropy between $p(\bm{s|\bm{y}})$ and $q_{\bm{\phi}}^{\prime \prime}(\bm{s}|\bm{y})$ becomes
\begin{align}    
      \! - \! \expct_{p(\! \bm{y} \! )} \!    \! \left[  \expct_{p(\! \bm{s}|\bm{y}\! )}\! \! \left[   \log   \!\left( q_{\bm{\phi}}^{\prime \prime}(\! \bm{s}|\bm{y} \!  ) \right) \! \right] \!  \right] \!  \!  = \!  \!   \expct_{p(\! \bm{y}\! )} \! \!  \left[    \expct_{p(\! \bm{s}|\bm{y}\! )}  \!    \left[  1 \! - \!  \text{SSIM}\! \left( \! \bm{s} ,\bm{\gamma}(\bm{y}) \! \right) \right]    \right] \!    , \! \label{SSIM_cross_entropy}
      \raisetag{5ex}
\end{align}
which we can minimize over $\bm{\phi}$, and thus over $\bm{\gamma}(\bm{y})$, by approximating the expected value over $\bm{y}$ and $\bm{s}$ using training data samples. This way we obtain the loss function of $1-\text{SSIM}(\bm{s}|\bm{y})$ 
to train the encoders and the decoders.



As before, using the maximum likelihood criterion to get the estimate $\hat{\bm{s}}$, we need to find the maximizer of $q_{\bm{\phi}}^{\prime \prime}(\bm{s}|\bm{y})$ with respect to $\bm{s}$, where its clear
from \eqref{decoder_SSIM} that is has to be in $[0,d]^L$.
Furthermore, the unique maximizer of $\text{SSIM}(\bm{s},\bm{\gamma}(\bm{y}))$ is $\bm{s}=\bm{\gamma}(\bm{y})$ \cite{hitchhikersguideSSIM}. 
As we have for $q_{\bm{\phi}}^{\prime \prime}(\bm{s}|\bm{y})$ a strictly monotonically increasing function of $\text{SSIM}(\bm{s},\bm{\gamma}(\bm{y}))$, its maximizer does not change.
Therefore, we have $\bm{s}=\bm{\gamma}(\bm{y})$ as the unique maximizer of $q_{\bm{\phi}}^{\prime \prime}(\bm{s}|\bm{y})$, which means our reconstruction estimate is $\hat{\bm{s}}=\bm{\gamma}(\bm{y})$.


\subsection{Combining MSE and SSIM Loss}
To solve \eqref{MinCrossEntropy}, we now assume the reconstruction decoder $q_{\bm{\phi}}(\bm{s}|\bm{y})$ to be a convex combination of the parameterized \acp{pdf} $q_{\bm{\phi}}^{\prime}(\bm{s}|\bm{y})$and $q_{\bm{\phi}}^{\prime \prime}(\bm{s}|\bm{y})$ with parameters $\bm{\mu}(\bm{y})$ and $\bm{\gamma}(\bm{y})$, respectively. 
Since we want a convex combination of \ac{MSE} and \ac{SSIM} as loss function, we do parameter sharing with $\bm{v}_{\bm{\phi}}(\bm{y}) \vcentcolon = \bm{\mu}(\bm{y})=\bm{\gamma}(\bm{y}) \in \mathbb{R}^L$, which depends on the \ac{NN} parameters $\bm{\phi}$. The combined \ac{pdf} can then be written as
\begin{equation}
    q_{\bm{\phi}}(\bm{s}|\bm{y}) =(1-\beta) \, q_{\bm{\phi}}^{\prime}(\bm{s}|\bm{y}) + \beta \,q_{\bm{\phi}}^{\prime \prime}(\bm{s}|\bm{y}),\label{q_phi_mix}
\end{equation}
with $\beta\in [0,1]$. The maximum likelihood estimate of the reconstruction is then given by  $\hat{\bm{s}}=\bm{v}_{\bm{\phi}}(\bm{y})$, as this is the maximizer of both $q_{\bm{\phi}}^{\prime}(\bm{s}|\bm{y})$ and $q_{\bm{\phi}}^{\prime \prime}(\bm{s}|\bm{y})$, as discussed above.


As we assumed the variance $\sigma^2$ of the Gaussian $q_{\bm{\phi}}^{\prime}(\bm{s}|\bm{y})$ to be fixed, we choose $\sigma^2=L/2$ here without restricting generality, as 
all cases of weighting \ac{MSE} and \ac{SSIM} are covered with $\beta\in[0,1]$.
Using \eqref{MSE_derived} and \eqref{SSIM_cross_entropy}, we can calculate the cross entropy  between \eqref{q_phi_mix} and $p(\bm{s}|\bm{y})$,
and by ignoring the constant terms, we get the solvable minimization problem

\vspace*{-1.2cm}
\begin{align} \label{MinMSESSIM}
    &\underset{
        \substack{
            \bm{\theta}_1,\dots,\bm{\theta}_N,\, \bm{\phi},\, \bm{\psi}
        }
    }{\argmin} \ 
    \begin{array}{l}
    \\  \ \
    \\ \ \
    \alpha \Bigl( \,\expct_{p(\bm{y})} \left[ \expct_{p(\bm{s}|\bm{y})} \left[ (1-\beta) \,
        \text{MSE} \left( \bm{s},\bm{v}_{\bm{\phi}}(\bm{y}) \right)
    \right. \right.  \Bigl.  \\
     \Bigl. \left. \left. \quad + \beta \left( 1 - \text{SSIM} \left(  \bm{s},\bm{v}_{\bm{\phi}}(\bm{y}) \right) \right)
    \right] \right]  \Bigl) \\
     + (1 - \alpha) \, \expct_{p(\bm{y})} \left[
        \entropy\left( p(\bm{z}|\bm{y}), q_{\bm{\psi}}(\bm{z}|\bm{y}) \right)
    \right]\end{array} \raisetag{27pt}  \\
    &\text{s.t.} \quad 
     \bm{c}_i \in \mathbb{R}^{N_{\text{Tx}}/N}, \quad
    P_i \leq 1, \quad \nonumber 
    \text{for } i = 1,\dots,N .
\end{align}

\section{Simulation Results}\label{sec:simulation_results}


We evaluate the proposed semantic communication system for data reconstruction and task execution for an example setting of image reconstruction and image classification using the CIFAR-10 dataset consisting of color images of size  $32\times32\times3$  \cite{Krizhevsky09learningmultiple}. 
We consider a multi-view scenario with $N=4$ sensing nodes, where each node has access to a non-overlapping square quarter of the image, corresponding to an image of size $16\times16\times3$.
Each sensing node independently encodes the data and transmits over independent channels to a central receiver for classification and image reconstruction.

We use the ResNet14 architecture \cite{7780459} for the encoders and reconstruction decoder. Each encoder consists of one convolution layer, six residual blocks, and a fully connected layer with power normalization. The reconstruction decoder includes two fully connected layers followed by residual blocks (one standard, two transposed for upsampling, and one standard) and a final transposed convolution layer. The classification decoder consists of three fully connected layers\footnote{The code is available at https://github.com/ant-uni-bremen/Semantic-Communication-for-Task-Execution-and-Data-Reconstruction}. 


We train the \ac{NN} parameters of the proposed semantic communication system in an end-to-end manner according to \eqref{MinMSESSIM} for $\alpha\in[0,1]$ and $\beta\in[0,1]$. We use a batch size of $32$ and a learning rate of $10^{-4}$.
For the cases of $\alpha=0$ and $\alpha=1$, the respective objective of data reconstruction and task execution in \eqref{MinMSESSIM} vanishes, meaning that the respective decoder is not trained at all.
To mitigate this problem, we first train the whole system for $300$ epochs. Then the encoder weights are frozen and the two decoders are trained for another $300$ epochs, which are independent of $\alpha$, as discussed before for \eqref{enc_dec_objective}. 
This training procedure is shown in Fig. \ref{fig:training_staility} for the image classification accuracy over the training epochs for some combinations of $\alpha$ and $\beta$ values. 
For $\alpha=1$, the classification accuracy is $0.1$ for the first $300$ epochs, as the classification decoder is only trained in the second training phase. For $\alpha <1$, the classification decoder is trained normally, and the training converges rather quickly after about $100$ epochs.


\begin{figure}[t]
    \centering
    \scalebox{0.85}{\input{tikz/results/alpha_beta_training_stability.tikz}}
    \vspace{-1em}
    \caption{
    Task execution error rate over training epochs for different $\alpha$ and $\beta$ values.
    }
    \vspace{-1.8em} 
    \label{fig:training_staility}
\end{figure}


For the following simulations, the number of channel uses per encoder is $N_\text{Tx}/N=50$, our channel model is an \ac{AWGN} channel, and the training and evaluation \ac{SNR} per channel use is $\SI{3}{\decibel}$.
We evaluate the task execution accuracy by the classification accuracy, i.e., the ratio of correctly classified images. The data reconstruction performance is measured by \ac{PSNR} and \ac{SSIM}. The \ac{PSNR} between reconstructed image $\hat{\bm{s}}$ and true image $\bm{s}$ with our normalization is given as 
\mbox{$\text{PSNR}=10\log_{10}\left(\frac{1}{ \frac{1}{L}\sum_{n=1}^L \! \left(s_n - \hat{s}_{n}\right)^2}\right)$}.
For $\text{SSIM}(\bm{s},\hat{\bm{s}})$ \eqref{ssim}, we use the standard TensorFlow implementation using Gaussian weighted windows \cite{hitchhikersguideSSIM}. For both \ac{PSNR} and \ac{SSIM}, larger values correspond to higher similarity between true and estimated image.

For comparison, two baseline methods are considered.
As a digital communication baseline, the images are subsampled, quantized and Huffman coded, followed by an LDPC channel code of rate $\tfrac{1}{2}$. 
The average number of channel uses per image is set to $N_\text{Tx}/N = 50$. The subsampling factor, the number of quantization bits, and the modulation order are optimized via grid search for image reconstruction at $\mathrm{SNR}=3,\text{dB}$.
When evaluated with a classifier trained independently on the CIFAR-10 dataset, classification of the reconstructed images effectively fails, achieving an accuracy only slightly above the random baseline of $0.1$.
As an upper bound on the accuracy of the multi-view classification problem, a perfect channel baseline is included with uncompressed and noiseless transmission.

\begin{figure}[t]
    \vspace{-0.6em}
    \centering
    \scalebox{0.9}{
\begin{tikzpicture}

\definecolor{beta1}{RGB}{0, 119, 187}  

\definecolor{beta2}{RGB}{51, 187, 238} 

\definecolor{beta3}{RGB}{0, 153, 136} 

\definecolor{beta4}{RGB}{238, 119, 51} 

\definecolor{beta5}{RGB}{204, 51, 17} 

\definecolor{darkslategray38}{RGB}{38,38,38}
\definecolor{lightgray204}{RGB}{204,204,204}

\begin{groupplot}[
group style={group size=1 by 3,
            vertical sep=0.10cm},
  grid=both,
  major grid style={color=lightgray204},
  minor grid style={dotted, color=lightgray204},
  axis line style={very thick, color=black},
  tick style={color=black, thick},
  tick label style={color=black},
  label style={color=black},
  width=8cm, height=3.2cm, 
  scale only axis,
]
\nextgroupplot[
ylabel={PSNR [dB]},
  ymin=11, ymax=22,
  xticklabels={},
  xtick={0,...,10},
  ytick={12,14,16,18,20,22},
  yticklabels={\raisebox{0pt}{12},14,16,18,20,\raisebox{-12pt}{22}}
]
\addplot [semithick, beta1, solid, mark=*, mark size=3, mark options={}]
table {%
0 13.7227592468262
1 13.7467641830444
2 16.0598678588867
3 16.7871532440186
4 17.3879451751709
5 17.6461944580078
6 18.0348072052002
7 18.5686473846436
8 20.0002326965332
9 20.3041896820068
10 20.27272605896
};
\addplot [semithick, beta2, solid, mark=*, mark size=3, mark options={solid}]
table {%
0 13.4153060913086
1 13.4712810516357
2 16.71201516357
3 17.6594524383545
4 18.1193466186523
5 18.2120208740234
6 18.5850410461426
7 18.8643321990967
8 19.4342212677002
9 19.45441674103
10 19.4886913299561
};
\addplot [semithick, beta3, solid, mark=*, mark size=3, mark options={solid}]
table {%
0 13.2686519622803
1 13.2505512237549
2 16.9342880249023
3 17.7221965789795
4 18.0426597595215
5 18.2376976013184
6 18.4860363006592
7 18.6342277526855
8 19.014533996582
9 19.0287208557129
10 19.0813617706299
};
\addplot [semithick, beta4, solid, mark=*, mark size=3, mark options={solid}]
table {%
0 13.06214427948
1 13.1363277435303
2 16.6644897460938
3 17.7326812744141
4 18.0468196868896
5 18.169900894165
6 18.3701152801514
7 18.4488086700439
8 18.7088165283203
9 18.7377147674561
10 18.7272281646729
};
\addplot [semithick, beta5, solid, mark=*, mark size=3, mark options={solid}]
table {%
0 12.6969547271729
1 12.735200881958
2 16.9001216888428
3 17.6086616516113
4 17.8367652893066
5 17.9349117279053
6 18.1554412841797
7 18.1962242126465
8 18.3728809356689
9 18.4279251098633
10 18.457498550415
};

\addplot [line width=1.5pt, black, dotted]
coordinates {(0,12.0162925720215) (10, 12.0162925720215)};
\node[] at (78,18) {baseline: digital comm.};

\nextgroupplot[
ylabel={SSIM},
  ymin=0.1, ymax=0.8,
  xticklabels={},
  xtick={0,...,10},
  ytick={0.1,0.2,0.3,0.4,0.5,0.6,0.7,0.8},
  yticklabels={\raisebox{8pt}{0.1},0.2,0.3,0.4,0.5,0.6,0.7,\raisebox{-12pt}{0.8}},
]
\addplot [semithick, beta1, solid, mark=*, mark size=3, mark options={solid}]
table {%
0 0.182491898536682
1 0.181702122092247
2 0.283788442611694
3 0.330943554639816
4 0.373282432556152
5 0.392693519592285
6 0.425419181585312
7 0.471704691648483
8 0.579501152038574
9 0.599465131759644
10 0.598529994487762
};
\addplot [semithick, beta2, solid, mark=*, mark size=3, mark options={solid}]
table {%
0 0.195941284298897
1 0.200366839766502
2 0.50815544474
3 0.555481255054474
4 0.598672270774841
5 0.609707593917847
6 0.641330063343048
7 0.665530562400818
8 0.702601671218872
9 0.70245746595
10 0.702257573604584
};
\addplot [semithick, beta3, solid, mark=*, mark size=3, mark options={solid}]
table {%
0 0.200005829334259
1 0.198901638388634
2 0.533423662185669
3 0.600096702575684
4 0.628497779369354
5 0.645395040512085
6 0.667406737804413
7 0.687316298484802
8 0.712564170360565
9 0.713358342647552
10 0.712643444538116
};
\addplot [semithick, beta4, solid, mark=*, mark size=3, mark options={solid}]
table {%
0 0.198300465941429
1 0.206293851137161
2 0.531485080718994
3 0.625052034854889
4 0.648823201656342
5 0.658731877803802
6 0.681584715843201
7 0.698677659034729
8 0.717558622360229
9 0.717574536800385
10 0.71525514125824
};
\addplot [semithick, beta5, solid, mark=*, mark size=3, mark options={solid}]
table {%
0 0.197740510106087
1 0.20434932410717
2 0.579805076122284
3 0.634976446628571
4 0.657059490680695
5 0.66684091091156
6 0.688932597637177
7 0.704427003860474
8 0.719218313694
9 0.719315886497498
10 0.719749808311462
};
\addplot [line width=1.5pt, black, dotted]
coordinates {(0,0.257653951644897) (10, 0.257653951644897)};
\node[] at (78,210) {baseline: digital comm.};

\nextgroupplot[
xlabel={ \LARGE$\alpha$},
  ylabel={Classification accuracy},
  ymin=0.4, ymax=0.8,
  xtick={0,...,10},
  ytick={0.4,0.5,0.6,0.7,0.8},
  yticklabels={\raisebox{6pt}{0.4},0.5,0.6,0.7,\raisebox{-12pt}{0.8}},
  xticklabels={0,0.001, 0.25, 0.5, 0.75, 0.8, 0.9, 0.95, 0.99, 0.999,1},
  xticklabel style={rotate=70},
  xlabel style={yshift=-2.5mm}, 
  legend cell align={left},
  legend style={
    draw=black,
    fill=white,
    fill opacity=1,
    font=\footnotesize,
    at={(0.28,0.55)}, 
    anchor=north,
    legend columns=2, 
  }
]
\addplot [semithick, beta1, solid, mark=*, mark size=3, mark options={solid}]
table {%
0 0.7022
1 0.7045
2 0.7096
3 0.7107
4 0.7104
5 0.7048
6 0.701
7 0.6961
8 0.6137
9 0.5145
10 0.5201
};
\addlegendentry{\small $\beta=0$}
\addplot [semithick, beta2, solid, mark=*, mark size=3, mark options={solid}]
table {%
0 0.7124
1 0.7095
2 0.7074
3 0.7104
4 0.7015
5 0.6987
6 0.6793
7 0.6587
8 0.5269
9 0.487
10 0.4944
};
\addlegendentry{\small $\beta=0.25$}
\addplot [semithick, beta3, solid, mark=*, mark size=3, mark options={solid}]
table {%
0 0.7129
1 0.716
2 0.7133
3 0.6819
4 0.6843
5 0.6831
6 0.6761
7 0.6404
8 0.5202
9 0.4876
10 0.4861
};
\addlegendentry{\small $\beta=0.5$}
\addplot [semithick, beta4, solid, mark=*, mark size=3, mark options={solid}]
table {%
0 0.7038
1 0.7048
2 0.6955
3 0.6915
4 0.6729
5 0.6811
6 0.6552
7 0.6389
8 0.5068
9 0.4775
10 0.4796
};
\addlegendentry{\small $\beta=0.75$}
\addplot [semithick, beta5, solid, mark=*, mark size=3, mark options={solid}]
table {%
0 0.7154
1 0.7073
2 0.6961
3 0.6779
4 0.6688
5 0.6671
6 0.6579
7 0.6316
8 0.4986
9 0.4845
10 0.4828
};
\addlegendentry{\small $\beta=1$}

\addplot [line width=1.5pt, black, dashed]
coordinates {(0,0.7225) (10, 0.7225)};
\node[] at (78,342) {baseline: perfect channel};

coordinates {(0,0.45) (10, 0.45)};

\end{groupplot}
\end{tikzpicture}}
    \vspace{-0.2em}
    \caption{
    Trade-off between \ac{PSNR},  \ac{SSIM}, and classification accuracy across $\alpha$ values for different $\beta$ weighting \ac{MSE} and \ac{SSIM} loss, with $N \!= \!4$ sensing nodes, $N_{\text{Tx}}/N \!=  \!50$ channel uses per sensing node, and $\text{SNR} \! = \! \SI{3}{\decibel}$.
    }
    \vspace{-1em}
    \label{fig:ssim-psnr}
\end{figure}


We investigate the trade-off between data reconstruction and task execution over $\alpha$ for different weights $\beta$ between \ac{MSE} and \ac{SSIM}, with the results shown in Fig.~\ref{fig:ssim-psnr}.

First, we analyze the case where only the \ac{MSE} loss is used for image reconstruction ($\beta=0$), over different $\alpha$ values.
For small $\alpha$ values, the model favors classification accuracy and neglects transmitting features required for image transmission, achieving only a \ac{PSNR} of about $\SI{14}{\decibel}$ and an \ac{SSIM} of about $0.2$. When $\alpha$ is increased up to about $0.9$, the \ac{PSNR} is increased  to about $\SI{18}{\decibel}$ and \ac{SSIM} to about $0.42$, while the task accuracy remains basically stable. This indicates that for larger $\alpha$ up to $\alpha \approx 0.9$, features are selected for transmission that contain more information required for image reconstruction without having to sacrifice transmitting discriminative features required for classification.
For larger $\alpha$ beyond this point, the classification accuracy declines while the reconstruction performance continues to improve. For $\alpha=1$, the classification accuracy drops to about $0.5$, as mostly features are transmitted for fine visual details, which do not help much for classification. 


Furthermore, we investigate the trade-off between \ac{MSE} and \ac{SSIM}, where a larger $\beta$ means a larger weight on \ac{SSIM}.
This is reflected in the results, where larger $\beta$ increases \ac{SSIM}, but decreases \ac{MSE} for large values of $\alpha$.
It can be seen that $\beta>0$ leads to larger \ac{PSNR} in the range of $0.25\leq \alpha \leq 0.9$, which can be explained by the fact that $1-\text{SSIM}(\bm{s},\hat{\bm{s}})$ is larger than $\text{MSE}(\bm{s},\hat{\bm{s}})$, which leads to a larger weight for image reconstruction.
Furthermore, \ac{MSE} prioritizes pixel accuracy, preserving texture, but often causing blur, whereas \ac{SSIM} prioritizes structure, enhancing shape preservation.
A balanced trade-off between \ac{PSNR} and \ac{SSIM} and classification accuracy can be achieved for $\beta=0.25$ and $\alpha=0.75$, where the classification accuracy is $0.7$, $\text{PSNR}=\SI{18}{\decibel}$, and $\text{SSIM}=0.6$.
In comparison, for $\beta=0$ and $\alpha=0.9$, the classification accuracy remains $0.7$ and $\text{PSNR}=\SI{18}{\decibel}$, while the \ac{SSIM} decreases significantly to $\text{SSIM}=0.42$.

\section{Conclusion} \label{section.Conclusion}
We investigated a semantic communication system designed for joint task execution and data reconstruction. As both objectives cannot be optimally fulfilled at the same time, a trade-off has to be made. We modeled this trade-off by formulating a joint objective as a convex combination of both objectives of task execution and data reconstruction.

Our findings show that data reconstruction quality can be significantly improved without limiting task execution significantly for moderate weighting of data reconstruction with \mbox{$\alpha\leq 0.9$}.
This can be achieved by learning to transmit features that preserve fine visual detail for data reconstruction while still being as discriminative as possible for task execution.
\mbox{Finally}, it was shown under which assumptions we get the \ac{SSIM} loss from the mutual information maximization problem.







\bibliographystyle{IEEEtran}
\bibliography{IEEEabrv,References.bib}

\end{document}